\documentclass[12pt]{iopart}
\usepackage[lmargin=2.5cm,rmargin=2.5cm,tmargin=3cm,bmargin=2cm]{geometry}
\usepackage{setspace} 
\usepackage{hyperref}
\hypersetup{
    colorlinks=true,
    linkcolor=blue,
    filecolor=blue,      
    urlcolor=blue,
    citecolor=blue
}
\usepackage{graphicx}
\usepackage[font=small]{caption}
\usepackage[font=small]{subcaption}
\usepackage{hhline}

\usepackage{url}
\usepackage{footmisc}
\usepackage{xcolor}
\usepackage{etoolbox}
\usepackage{titleps}

\usepackage{cite}

\usepackage{tabularx} 
\usepackage{verbatim} 
\usepackage[capitalize]{cleveref}
\usepackage{xspace}

\newcommand{\SI}[2]{#1~\mathrm{#2}}

\newcommand{\U}[1]{$^{#1}\mathrm{U}$\xspace}
\newcommand{\ufc}{\U{238}~FC\xspace}
\newcommand{\ufcs}{\U{238}~FCs\xspace}
\newcommand{\ddng}{DD~NG\xspace}
\newcommand{\dtng}{DT~NG\xspace}
\newcommand{\dlos}{DLOS\xspace}
\newcommand{\bfield}{B-field\xspace}

\newcommand{\add}[1]{\textcolor{red}{#1}}
\renewcommand{\add}[1]{#1\unskip}

\usepackage{graphicx} 

\begin{document}

    \title[]{Characterization of a prototype parallel-plate \U{238}~fission chamber with DD and DT fusion neutron~sources}
        \newcommand{\iPSFC}{$^1$\xspace}
    \newcommand{\iCFS}{$^2$\xspace}
    \newcommand{\iCCU}{$^3$\xspace}

    \newcommand{\PSFC}{\iPSFC Plasma Science and Fusion Center, Massachusetts Institute of Technology, Cambridge, MA, USA\xspace}
    \newcommand{\CFS}{\iCFS Commonwealth Fusion Systems, Devens, MA, USA\xspace}
    \newcommand{\CCU}{\iCCU Coastal Carolina University, Conway, SC, USA\xspace}

    \author{V.~Hagenlocker\iPSFC\footnote[1]{Shared first authorship},
            R.A.~Tinguely\iPSFC\footnotemark[1]\footnote[2]{Corresponding author: tinguely@psfc.mit.edu},
            X.~Wang\iPSFC\iCCU,
            J.L.~Ball\iPSFC, 
            B.~Buschmann\iPSFC,
            M.~Gatu-Johnson\iPSFC,
            R.~Gocht\iCFS, 
            P.~Raj\iCFS
            }
    
    \address{   \PSFC \\
                \CFS \\
                \CCU \\
            }
    \begin{abstract}
    The SPARC tokamak will employ \U{238}-based fission chambers \add{(FCs)} to monitor high-performance deuterium-tritium (DT) plasma operations, spanning neutron yield rates from ${\sim}10^{15}$ to ${>}10^{19}$ n/s. This work validates the \ufc design, which utilizes a parallel-plate detector geometry and borated polyethylene collimation to prioritize unscattered DD and DT fusion neutrons. Experimental testing with both DD and DT neutron generators corroborates vendor-specified efficiencies and demonstrates excellent detector linearity, with measured count rates showing good agreement with OpenMC neutronics simulations. Further characterization confirms the \ufc's robustness against SPARC-relevant environmental challenges, including stray magnetic fields up to $\SI{14}{mT}$ and possible signal degradation risks associated with ${\sim}\SI{30}{m}$ long cable runs. These results confirm that the \ufc, supported by indirect neutron shielding as well as appropriate pulse height thresholds, provides a reliable solution for fusion power measurements as part of the SPARC neutron diagnostics suite.
\end{abstract}
    
    \section{Introduction}\label{sec:intro}

\begin{figure}[h!]
        \centering
        \begin{subfigure}{0.49\linewidth}
            \centering
            \includegraphics[width=\linewidth]{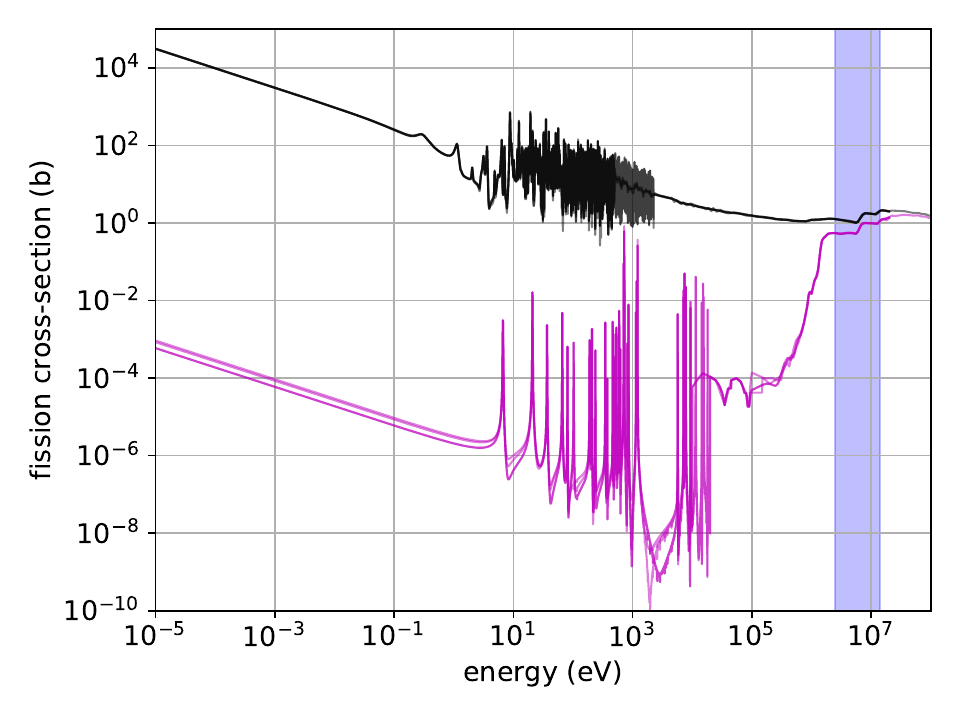}
            \caption{}
            \label{fig:cx}
        \end{subfigure}
        \begin{subfigure}{0.50\linewidth}
            \centering
            \includegraphics[width=\linewidth]{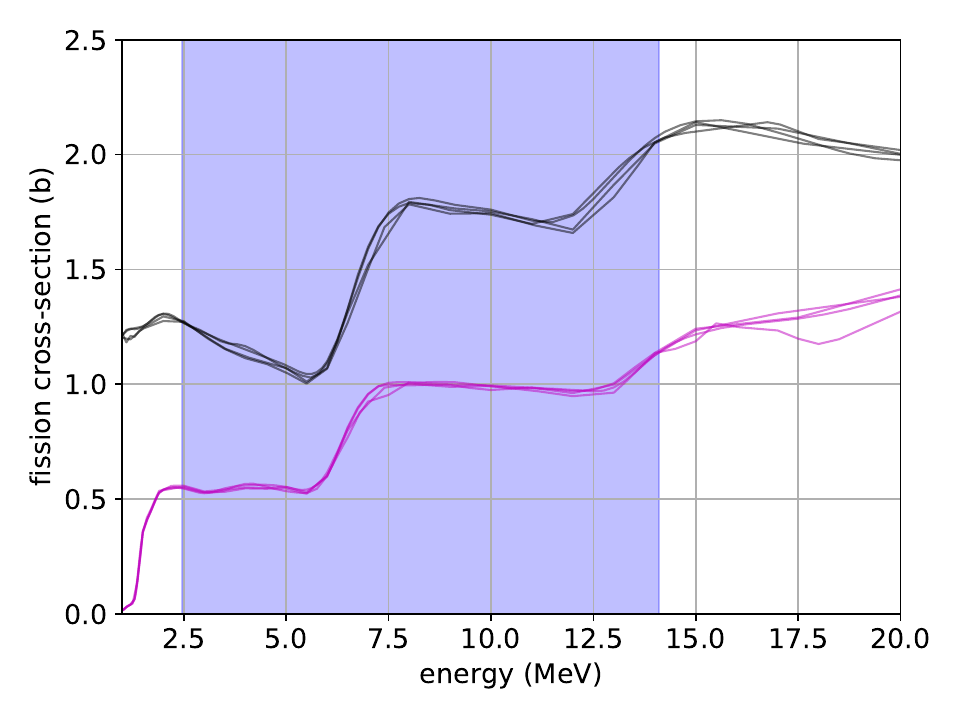}
            \caption{}
            \label{fig:zoom}
        \end{subfigure}
        \caption{Fission cross-sections for \U{235} (top, black) and \U{238} (magenta, bottom) from JENDL libraries 3.2, 3.3, 4.0, and 5 (300~K) \cite{JENDL32,JENDL33,JENDL4,JENDL5} for (a)~thermal to fast neutron energies and (b)~zoomed in on ${>}\SI{1}{MeV}$. The range for DDn and DTn birth energies ($\SI{2.45-14.1}{MeV}$) is shaded blue. Note the different vertical and horizontal axes.}
        \label{fig:cx_zoom}
    \end{figure}

    In many nuclear fusion experiments, nuclear fission is relied upon to monitor neutron production which correlates with plasma performance and experimental progress. The deuterium (D) and tritium (T) fusion reactions are often of most interest, with neutron energies $\SI{2.45}{MeV}$ and $\SI{14.1}{MeV}$ for DD and DT fusion, respectively. Uranium isotopes \U{235} and \U{238} are two common choices for the sensitive materials inside neutron detectors, or fission chambers (FCs) \cite{Jarvis1994}. Their total neutron-induced fission cross-sections are shown in \cref{fig:cx_zoom}. Data are plotted from JENDL libraries 3.2, 3.3, 4.0, and 5 (at 300~K) \cite{JENDL32,JENDL33,JENDL4,JENDL5} to highlight some uncertainty in the cross-sections. It is clear in \cref{fig:cx} that \U{235} (in black) is far more likely (by a factor of ${\sim}10^4$) to interact with thermal neutrons (${<}\SI{1}{eV}$) than fast neutrons (${>}\SI{1}{MeV}$), while the opposite is true for \U{238} (in magenta). Furthermore, from \cref{fig:zoom}, we see that \U{238} is ${\sim}2$ times more sensitive to DT than DD neutron energies.
    
    
    Within magnetic confinement fusion, many tokamaks have used uranium-based fission chambers to monitor plasma performance; these include Alcator C-Mod \cite{Fiore1995}, JET \cite{Jarvis1990,Batistoni2018}, TFTR \cite{Nieschmidt1985}, and others. By its nature of being a ``slow'' neutron detector, a \U{235}~FC requires moderation to thermalize ``fast'' fusion neutrons; \ufcs can directly measure these fast neutrons, but with lower efficiency. Therefore, the SPARC tokamak \cite{Creely2020}, currently being assembled by Commonwealth Fusion Systems, will utilize \ufcs as part of its neutron flux monitor suite \cite{Raj2024} for the highest DT fusion power operations (up to $\SI{140}{MW} \sim \SI{5\times10^{19}}{n/s}$). 

    While SPARC plans to perform an in-situ calibration with well-characterized neutron sources, only the most sensitive detectors \cite{Wang2026_BF3} will achieve sufficient counting statistics; the \ufcs will instead rely on alternative methods, e.g. cross-calibrations during plasma operations. Design decisions must therefore be made based on vendor specifications and experimental validation, motivating the work reported in this paper. For the prototype of the \ufc model chosen for SPARC, we characterize its sensitivity and linearity to both DD and DT neutrons; its response to externally applied magnetic fields; the effectiveness of neutron collimation and shielding; and optionality in the backend electronics chain. 
    

    \add{The rest of the paper is organized as follows: The experimental setup is described in \cref{sec:setup}, and experimental results are shown in \cref{sec:expt}, along with complementary neutronics simulations. A summary is then provided in \cref{sec:summary}.}

    \section{Experimental setup}\label{sec:setup}

    \subsection{\U{238} fission chamber}\label{sec:setup-fc}


        The \ufc tested is a model PFC338/450/U238 from \add{Exosens}.%
            \footnote{\add{The PFC338/450/U238 was developed by Centronic Ltd. (469940, England), a UK-based manufacturer of radiation detectors, unified under the Exosens brand in 2025.}}
        It has a unique parallel plate structure within its cylindrical housing, as seen in \cref{fig:U238_diagram_openmc}, engineered to most efficiently detect collimated neutrons. The outer \add{stainless steel} housing has a length of \add{${\sim}\SI{18}{cm}$} and diameter of ${\sim}\SI{7.5}{cm}$. \add{In total,} approximately $\SI{150}{mg}$ of \add{triuranium octoxide ($\mathrm{U_3 O_8}$, 99.98\% \U{238})} is coated on the electrodes\add{, each with a sensitive diameter ${\sim}\SI{3.3}{cm}$ and areal coating density ${\sim}\SI{450}{\mu g/cm^2}$}, contained within a pressurized gas at \add{a few atm}.
        As neutrons cause fission events, charged products carrying the released energy (${\sim}\SI{200}{MeV}$) ionize the gas and allow electrons to flow from anode to cathode. The vendor-supplied efficiency is $\SI{2.264 \times 10^{-4}}{cps/nv}$, with count rate cps = counts per second and the flux $\mathrm{nv}$ in units of $\mathrm{n/cm^2/s}$.%
            \footnote{Note that available specification sheets from \add{Exosens} list the \ufc efficiency as high as $\SI{5.5 \times 10^{-4}}{cps/nv}$ for $\SI{14}{MeV}$ neutrons \add{at an operating voltage $+\SI{500}{V}$}.} 
        The FC can be operated in current or pulse counting mode; due to the relatively low neutron counting efficiency, only the latter is explored here.  

        \newcommand{\tempwidth}{0.49\linewidth}
        \newcommand{\tempheight}{3.5cm}
        \begin{figure}[h!]
            \centering
            \begin{subfigure}{0.66\linewidth}
                \centering
                \includegraphics[height=\tempheight]{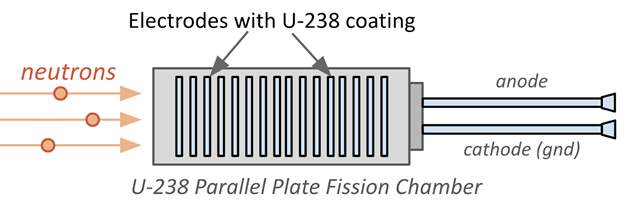}
                \caption{}
                \label{fig:U238_diagram}
            \end{subfigure}
            \begin{subfigure}{0.33\linewidth}
                \centering
                \includegraphics[height=\tempheight]{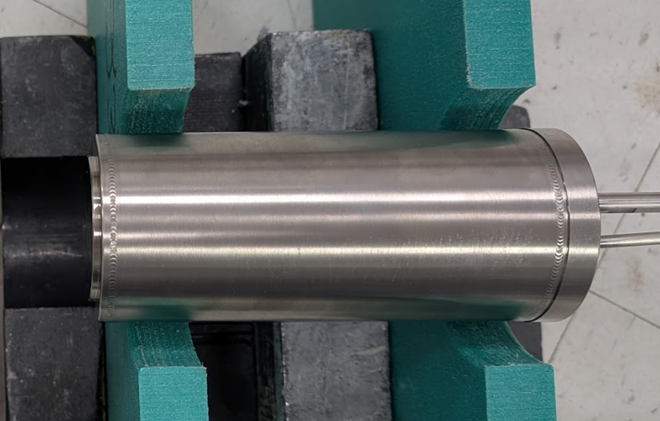}
                \caption{}
                \label{fig:U238}
            \end{subfigure}
            \caption{The parallel plate \U{238} fission chamber: (a)~schematic diagram of detector functionality, and (b)~top-down photograph, with plastic supports (green) braced by lead bricks.}
            \label{fig:U238_diagram_openmc}
        \end{figure}

    \subsection{Electronics and data acquisition}\label{sec:setup-daq}


        Two mineral insulated cable leads can be seen on the right in \cref{fig:U238,fig:U238_diagram}. Each inner conductor connects to one set of the ``interleaved'' parallel plates. In pulse counting mode, one lead is shorted to ground (by a grounding cap), while the other is connected to a high voltage power supply via a pre/amplifier. For all experiments detailed in \cref{sec:expt}, the applied high voltage is $+\SI{1}{kV}$,%
            \footnote{\add{According to the vendor, $+\SI{1}{kV}$ is the typical maximum operating voltage, although operation at higher voltages is possible.}}
        unless otherwise noted. Although not investigated in detail, operation at $+\SI{500}{V}$ does not appear to affect the detector performance\add{; future work should explore the sensitivity to operating voltage if lower values are required}.

        Two amplification options are tested: First is a pulse counter pre-amplifier, model CPA60 from Cooknell Electronics, Ltd., which outputs pulses with typical widths ${\sim}\SI{100}{ns}$. The second is an ORTEC model 140PC charge-sensitive preamplifier connected to an ORTEC model 590A amplifier, with pulse lengths ${\sim}\SI{10}{\mu s}$. Little difference is observed between the two options, which is - at least in part - likely due to the low count rate. Unless noted otherwise, the Cooknell preamplifier is the default choice for the experiments in \cref{sec:expt}.

        \begin{figure}[h!]
            \centering
            \includegraphics[width=0.5\linewidth]{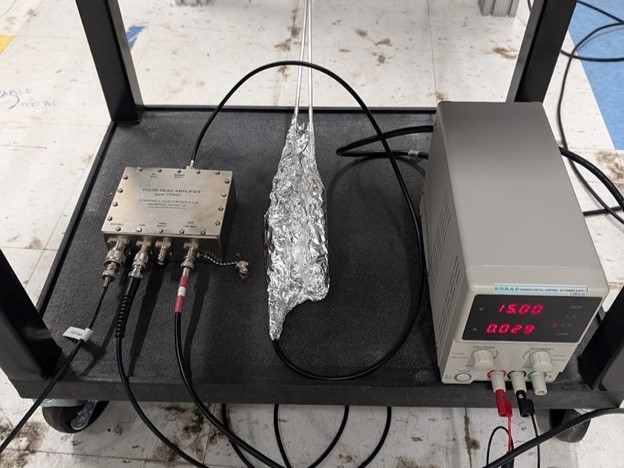}
            \caption{Left: Cooknell preamplifier for the U238 fission chamber (FC). Center: Leads of the \ufc wrapped in aluminum foil; one lead is grounded via a shorting cap (not visible). Right: Low voltage power supply for the preamplifier.}
            \label{fig:preamp_PS}
        \end{figure}

        The preamplifier itself is shown in \cref{fig:preamp_PS} (left), next to the \ufc leads (center) and the preamplifier's low voltage power supply (right). Note that the lead connections (including the grounding cap) are wrapped in aluminum foil. This is found to be needed in order to suppress high frequency oscillations in the detector signal, likely the result of some kind of electromagnetic pick-up or perhaps impedance mismatching, the cause of which is not entirely clear \add{to the authors}.

        Lastly, the signal is acquired by a 500~MSps, 14-bit CAEN model DT5730 digitizer. Such a high sampling rate is not really necessary for a neutron counter, especially since the \ufc does not measure neutron energy. However, this digitizer does allow us to investigate the pulse height spectrum (PHS, of fission product energy, effectively) and thresholds required for rejection of background signal and electronic noise. 

    \subsection{DD and DT neutron generators}\label{sec:setup-ng}


        Two accelerator-based neutron generators (NGs) are used in this study. The \ddng is a ThermoFisher Scientific model P385, while the \dtng is model A325; they are shown in \cref{fig:U238_DD,fig:U238_DT}, respectively. The nominal source rates are ${\sim}\SI{10^7}{DD~n/s}$ and ${\sim}\SI{10^8}{DT~n/s}$. \add{The \dtng's angular distribution of neutron energy spectra and flux has been characterized previously in \cite{TSakabe_2024}; the same characterization has not yet been completed for the \ddng.} In the experiments of \cref{sec:expt}, the \ufc is aligned at ${\sim}\SI{90}{deg}$ to the NG axis and in line with the target plane (as seen in \cref{fig:U238_DD_DT}), unless otherwise stated. This is done to ensure neutron energies as close as possible to birth energies in the center-of-mass frame.

        \renewcommand{\tempwidth}{0.49\linewidth}
        \begin{figure}[h!]
            \centering
            \begin{subfigure}{\tempwidth}
                \includegraphics[width=\textwidth]{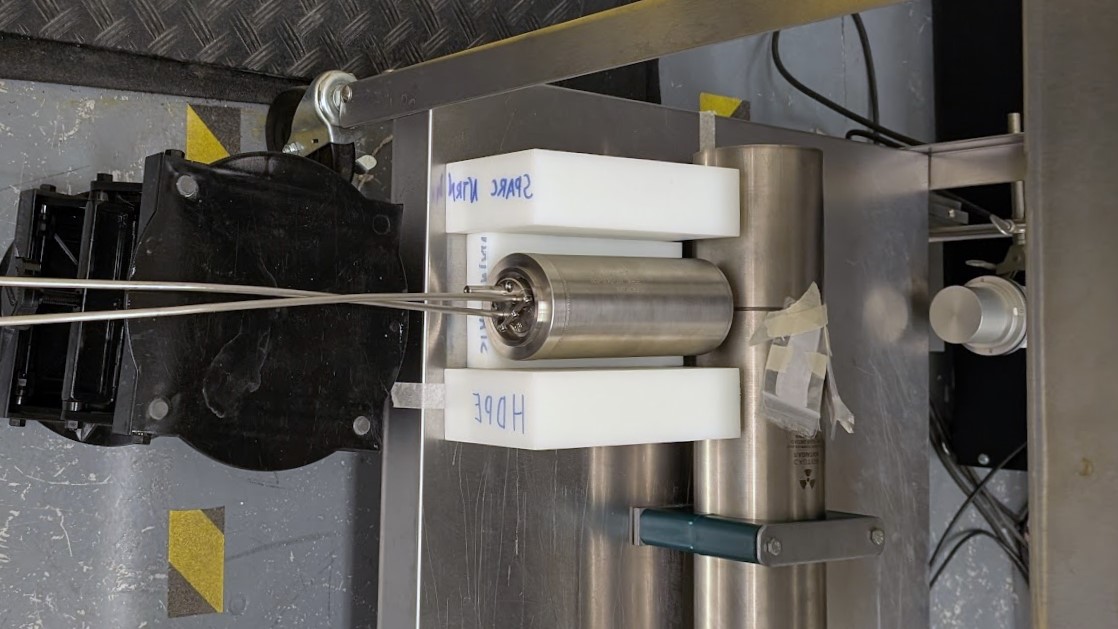}
                \caption{Left:\ufc. Center:\ddng. Right:DLOS.}
                \label{fig:U238_DD}
            \end{subfigure}
            \begin{subfigure}{\tempwidth}
                \includegraphics[width=\textwidth]{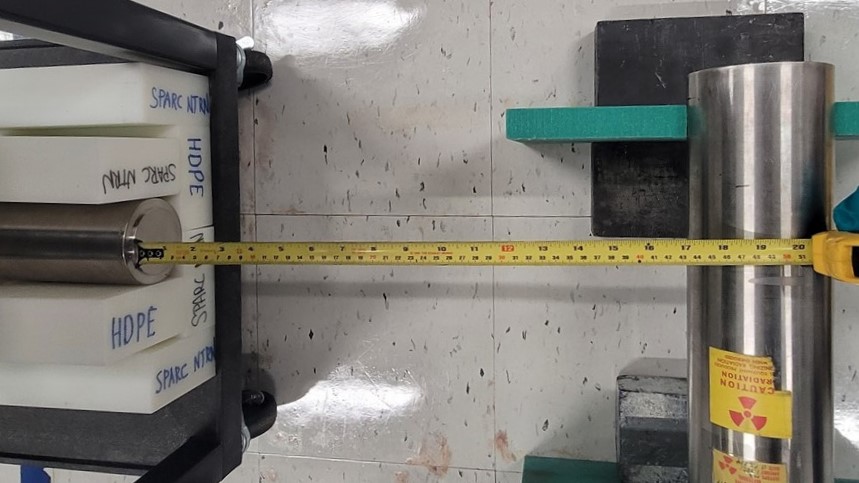}
                \caption{Left: \ufc. Right: \dtng.}
                \label{fig:U238_DT}
            \end{subfigure}
            \caption{The \ufc, supported by high density polyethylene (HDPE) blocks, placed within the radiation fields of (a)~DD and (b)~DT neutron generators (NG). In (a), the deuterated liquid organic scintillator (DLOS) reference detector can be seen on the right.}
            \label{fig:U238_DD_DT}
        \end{figure}

        Note that in \cref{fig:U238_DD_DT}, the \ufc is surrounded by white high density polyethylene (HDPE) bricks. Except for the test of neutron collimation and shielding in \cref{sec:expt-coll}, \add{these bricks are only used} to support the \ufc. No moderation is included between the \ufc and neutron source in order to prioritize direct neutron flux. While some energy-down-scattered neutrons will reach the \U{238}-coated plates, we expect the impact to be negligible given the steep drop in sensitivity below ${<}\SI{1}{MeV}$ (see \cref{fig:cx}).

    \subsection{Deuterated liquid organic scintillator reference detector}\label{sec:setup-dlos}


        To provide a high-count-rate reference measurement of the NG output rate, a 2-inch-by-2-inch cylindrical EJ-301D \cite{Becchetti2016} deuterated-xylene-based liquid organic scintillator (\dlos) is fielded. Manufactured by Eljen Technology, this detector consists of a 2-inch right cylindrical volume of scintillator coupled to a Hamamatsu R7724 photomultiplier tube. Unlike the \ufc, \dlos is capable of separating neutron vs gamma-induced pulses via pulse shape discrimination (PSD), which is used to isolate the neutron count rate for the analyses presented here. The excellent PSD performance of this specific detector has been demonstrated before in the literature; for example, see \cite{Ball2024}. 
        
        \add{Similar high voltages (${\sim}\SI{1}{kV}$) and nearly identical digitizer settings are used for the DLOS measurements of DD and DT neutrons below, although slightly altered PSD settings allow for improved neutron/gamma separation in the two different experiment environments. The inferred NG source rate is calculated in a similar way as the procedure detailed in \cite{Ball2026}, but instead using a simplified geometrical approximation for neutron transport from the point-like source, instead of Monte Carlo neutronics.}


    \section{Experimental characterization}\label{sec:expt}

    \subsection{Sensitivity and linearity}\label{sec:expt-lin}



\begin{figure}[h!]
            \centering
            \includegraphics[width=0.5\linewidth]{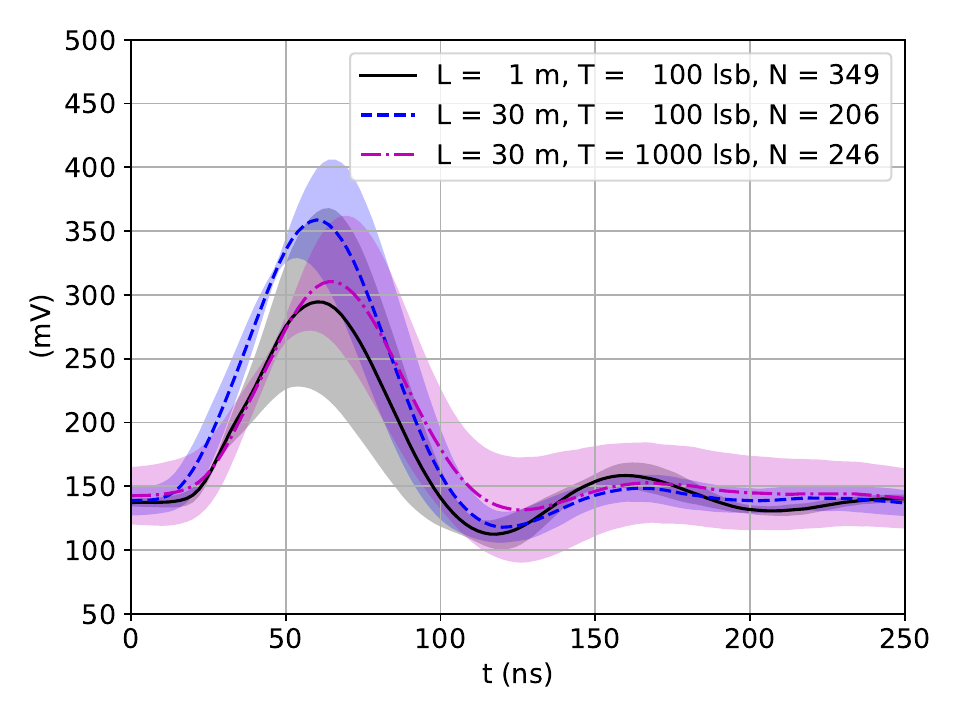}
            \caption{\add{Digitized voltage waveforms from the \ufc for two cable lengths $L = \SI{[1,30]}{m}$ and two thresholds $T = \SI{[100,1000]}{lsb}$. For $N$ waveform samples, the mean curve and shaded standard deviation is shown.}}
            \label{fig:wf}
        \end{figure}

        Before detector characterization, we first test our measurement sensitivity to the electronics and digitization chain. As discussed in \cref{sec:setup-daq}, the primary difference between preamplifiers is the pulse length, but not the shape of the pulse height spectrum (PHS) or total count rate. Typical signal waveforms and PHS from the \ufc are captured in \cref{fig:wf,fig:dd_dt_lin_phs}, respectively; here, the Cooknell preamplifier is used, as evidenced by the $O(\SI{100}{ns})$ pulse durations. In \cref{fig:wf}, we plot the ``mean'' waveform averaged over $N>200$ pulses, with standard deviations shown. For all experiments in the following sections, we use a signal cable of length $L = \SI{1}{m}$ between the \ufc and preamplifier; however, we see that using a $\SI{30}{m}$ long cable does not drastically change the waveforms. The digitization threshold - here in units of least significant bit (lsb) - can also be adjusted to filter out any electronic noise. Lastly, we note the oscillatory nature of the waveform beyond $t > \SI{100}{ns}$, which may be due to an impedance mismatch; however, the overall effect on the PHS and total count rate is minimal.

        Next, we assess the \ufc's linearity with the neutron source rate. Setups for the tests with the DD and DT NGs are shown in \cref{fig:U238_DD_DT}. Due to the \ddng's relatively low neutron source rate (${\sim}\SI{10^7}{n/s}$, see \cref{sec:setup-ng}) and the \ufc's low sensitivity, the FC is placed as close to target plane as possible, as seen in \cref{fig:U238_DD}, with ${\sim}\SI{8}{cm}$ of distance between the face of the FC and the center of the NG. The primary motivation for this is to achieve sufficient statistics $O(\SI{1000}{counts})$ in reasonable durations $O(\mathrm{hours})$. The \dlos is located on the opposite side of the \ddng, also in line with the target plane, as a stationary reference detector.

        A relatively high \add{digitization} threshold is set at $T = \SI{1000}{lsb}$ to reject low amplitude noise in this configuration, resulting in the PHS of \cref{fig:dd_lin_phs}. A scan in the \ddng accelerating voltage is performed from $\vert V \vert = \SI{100-130}{kV}$; as expected, the PHS grows with increasing (absolute) voltage, but the shapes are self similar. The \add{PHS} is integrated above a threshold (here, channel 64 of 1024) to calculate the \ufc's total count rate, which is plotted in \cref{fig:dd_lin} vs the total \ddng source rate as measured by \dlos (described in \cref{sec:setup-dlos}). Uncertainties due to counting statistics in the \ufc count rate are plotted but too small to see; uncertainties in the \dlos measurement are not provided, however, and could be as large as 100\%. Yet the order of magnitude, ${\sim}\SI{5\times 10^6}{DD~n/s}$, matches expectation and provides confidence in the DLOS measurement.

        \renewcommand{\tempwidth}{0.49\linewidth}
        \begin{figure}[h!]
            \centering
            \begin{subfigure}{\tempwidth}
                \centering
                \includegraphics[width=\linewidth]{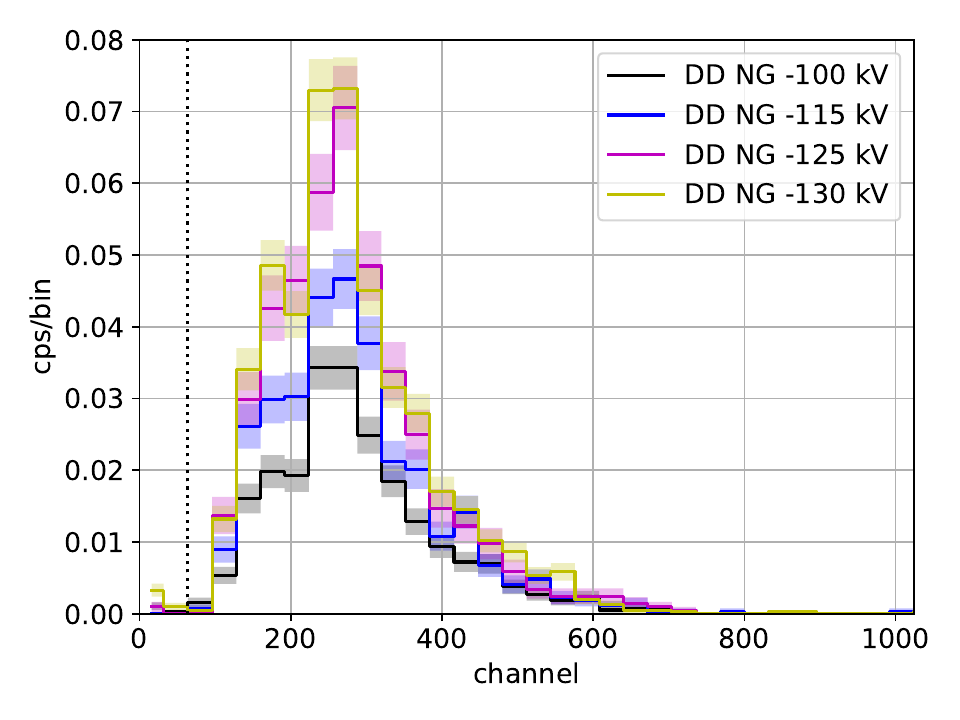}
                \caption{}
                \label{fig:dd_lin_phs}
            \end{subfigure}
            \begin{subfigure}{\tempwidth}
                \centering
                \includegraphics[width=\linewidth]{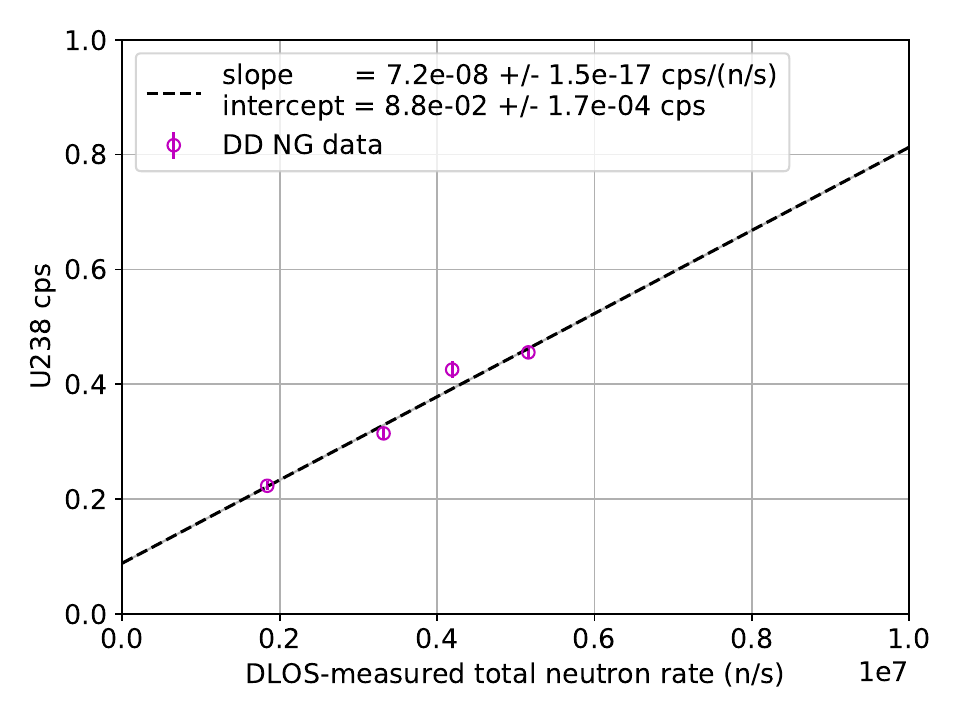}
                \caption{}
                \label{fig:dd_lin}
            \end{subfigure}
            \begin{subfigure}{\tempwidth}
                \centering
                \includegraphics[width=\linewidth]{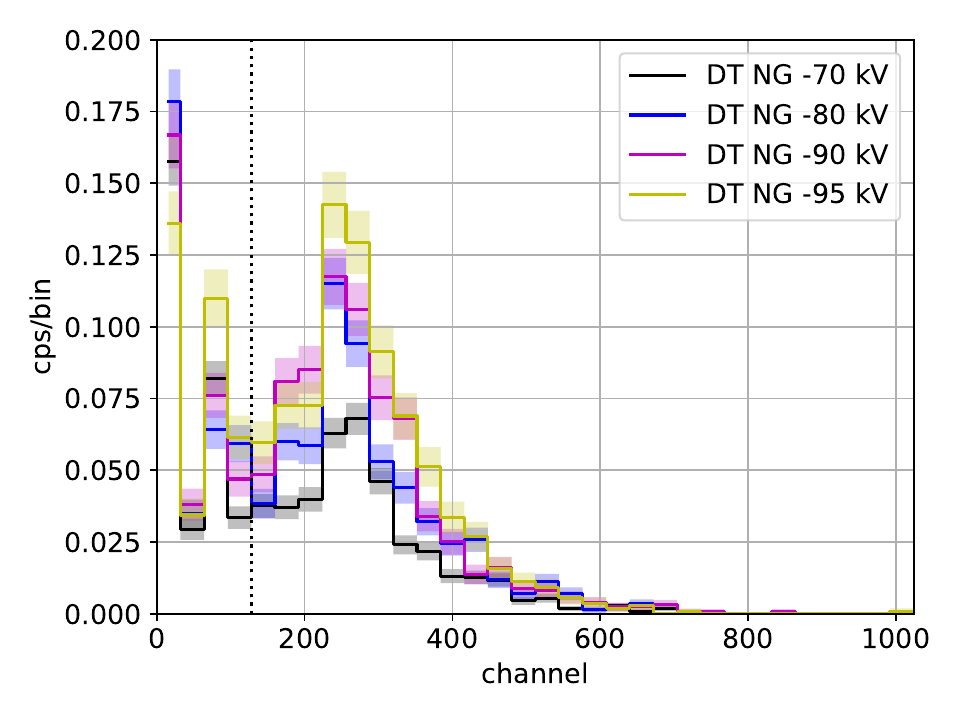}
                \caption{}
                \label{fig:dt_lin_phs}
            \end{subfigure}
            \begin{subfigure}{\tempwidth}
                \centering
                \includegraphics[width=\linewidth]{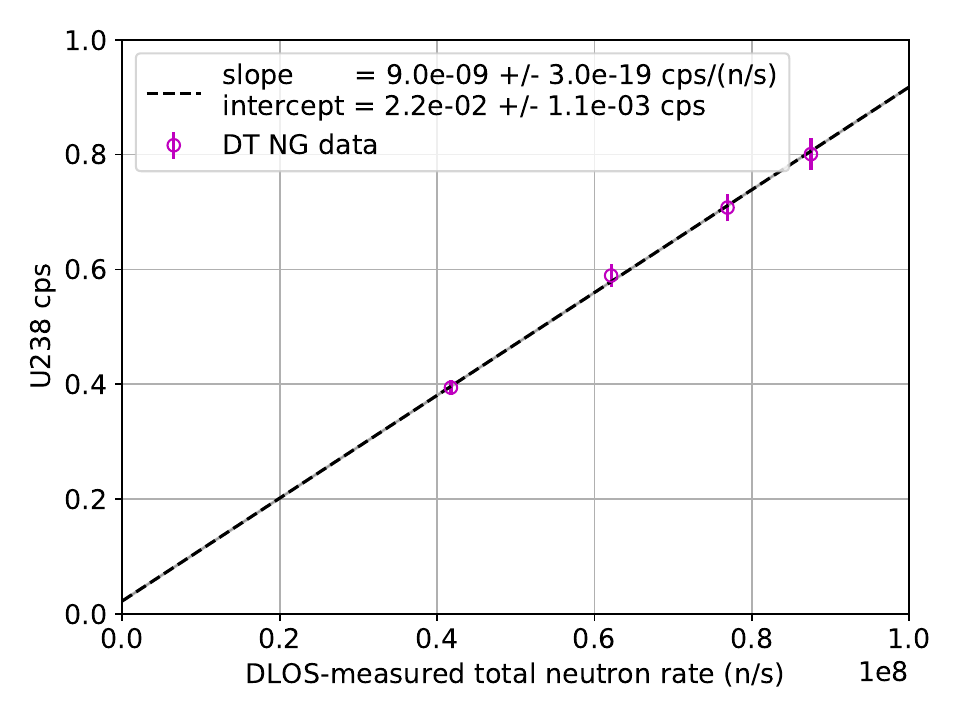}
                \caption{}
                \label{fig:dt_lin}
            \end{subfigure}
            \caption{(a,c)~Pulse height spectra and (b,d)~detector linearity for the \ufc within the radiation fields of the (a,b)~DD and (c,d)~DT neutron generator (NG). (a,c)~Accelerating voltages are listed for each NG, and the vertical dotted black line indicates the threshold. (b,d)~Data points integrated above the threshold vs the total NG yield rate as measured by a reference DLOS detector, with a linear fit shown in dashed black. Note the different horizontal and vertical axes.}
            \label{fig:dd_dt_lin_phs}
        \end{figure}
        
        A similar experiment is run using the \dtng, with the setup shown in \cref{fig:U238_DT}. Here, the face of the \ufc is ${\sim}\SI{50}{cm}$ from the center of the \dtng, which is possible due to the $10$ times increase in source rate compared to the \ddng. The accelerating voltage is ramped from $\vert V \vert = \SI{70-95}{kV}$. Complementary PHS and count rate data vs \dlos measurements are given in \cref{fig:dt_lin_phs,fig:dt_lin}, respectively. A lower \add{digitization} threshold is possible in this configuration, meaning that additional features of the PHS can be seen in \cref{fig:dt_lin_phs}: the lowest bin is likely noise, while the peak centered at channel~$\sim$80 is typically attributed to alpha decay; the data above this channel (and seen in \cref{fig:dd_lin_phs}) are due to the other fission fragments \cite{Taieb2016}. 

        A higher \add{PHS} threshold (here, channel~128 of 1024) is applied to calculate the total count rate, to be consistent with the \ddng measurement; however, we note that the result does not change significantly when lowering the threshold to include the ``alpha peak''. As with the \ddng, the \dlos measurement of the \dtng source rate is the expected order of magnitude $O(\SI{10^8}{DT~n/s})$. For both DD and DT data, the \ufc response shows impressive linearity with the neutron rate. The linear fit's extrapolation to $\SI{0}{(n/s)}$ is quite low for the \dtng (${\sim}\SI{0.02}{cps}$), which is promising. The same value for the \ddng (${\sim}\SI{0.09}{cps}$) is higher than expected - or desired - for future operation. Perhaps this offset is due to the higher threshold and exclusion of some data in this experimental configuration.

        Finally, we estimate the \ufc efficiency and DD vs DT neutron sensitivity. As mentioned, the vendor-specified efficiency is $\epsilon \approx \SI{2.3 \times 10^{-4}}{cps/nv}$. The \dtng experimental setup is better suited for this estimate given the ${\sim}\SI{50}{cm}$ separation between the \dtng and \ufc. While not strictly true for an accelerator-based neutron source, let us assume that the neutrons are emitted isotropically into $\SI{4\pi}{steradians}$; the ``effective'' flux evaluated at the center of the \ufc \add{sensitive volume} is then calculated applying a factor of $1/4\pi r^2$ with \add{$r \approx \SI{50 + 6.5 = 56.5}{cm}$}. Using the slope evaluated in \cref{fig:dt_lin}, we calculate an efficiency for DT neutrons: \add{$\epsilon_\mathrm{DT} \approx \SI{3.6\times10^{-4}}{cps/(n/cm^2/s)}$}. 
        
        A similar procedure is followed for the \ddng data, although the closeness of the \ddng and \ufc invalidates the small solid angle assumption. \add{In addition, the spatial anisotropy of NG emission would play an even larger role here, but is again not included.} Nevertheless, using \add{$r \approx \SI{8 + 6.5 = 14.5}{cm}$} and the slope from \cref{fig:dd_lin}, we find a DD neutron efficiency of \add{$\epsilon_\mathrm{DD} \approx \SI{1.9\times10^{-4}}{cps/(n/cm^2/s)}$}. Even though these are rough estimates, they are at least consistent with the vendor's values and with the expected increase in sensitivity from DD to DT neutron energies, i.e. from the \U{238} fission cross-sections (see \cref{fig:zoom}).

        Note that the \ufc's gamma sensitivity could not be investigated in this study due to (i)~the low emission rates of available gamma sources, as well as (ii)~the detector's expected low sensitivity to gammas. While high energy photons, such as n-gammas, can interact with the ionizing gas, their energy content $O(\SI{1}{MeV})$ is anticipated to be much lower than the fission products $O(\SI{100}{MeV})$, meaning gammas would register with relatively low pulse heights and could be filtered out of the \ufc data. While further analysis is needed, preliminary OpenMC simulations of neutron-induced prompt gamma emission \cite{Panontin2026} suggest that n-gamma fluxes are reduced by roughly an order of magnitude compared to neutrons, with energies ${<}\SI{10}{MeV}$, \add{indicating that SPARC's \ufcs should be robustly insensitive to the gamma background}. 

    \subsection{Collimation and shielding}\label{sec:expt-coll}


        As discussed in \cref{sec:setup-fc}, the parallel-plate design of this \ufc is tailored to maximize \add{the detection} efficiency of collimated neutrons, i.e. those with trajectories normal to the plates and along the detector's cylindrical axis. Collimation \add{for SPARC's \ufcs} will be achieved using a 5\%-borated high density polyethylene shielding material and a \add{cylindrical collimator}  of diameter $\SI{3.3}{cm}$ and variable length; see \cite{Raj2024,Raj2026} for more details. 

        \renewcommand{\tempwidth}{0.32\linewidth}
        \renewcommand{\tempheight}{6cm}
        \begin{figure}[h!]
            \centering
            \begin{subfigure}{0.4\linewidth}
                \centering
                \includegraphics[height=\tempheight]{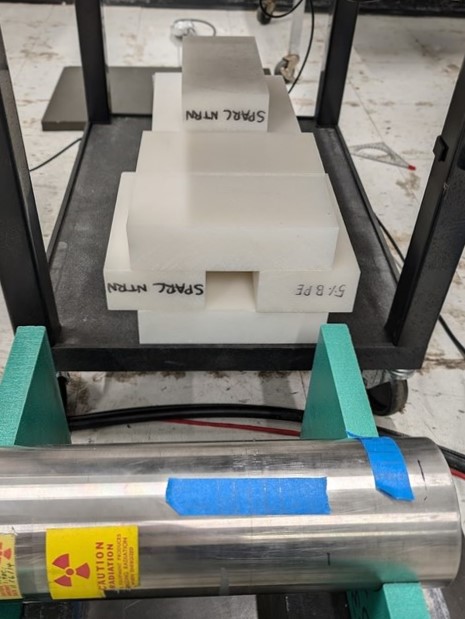}
                \caption{}
                \label{fig:collimator}
            \end{subfigure}
            \begin{subfigure}{0.59\linewidth}
                \centering
                \includegraphics[height=\tempheight]{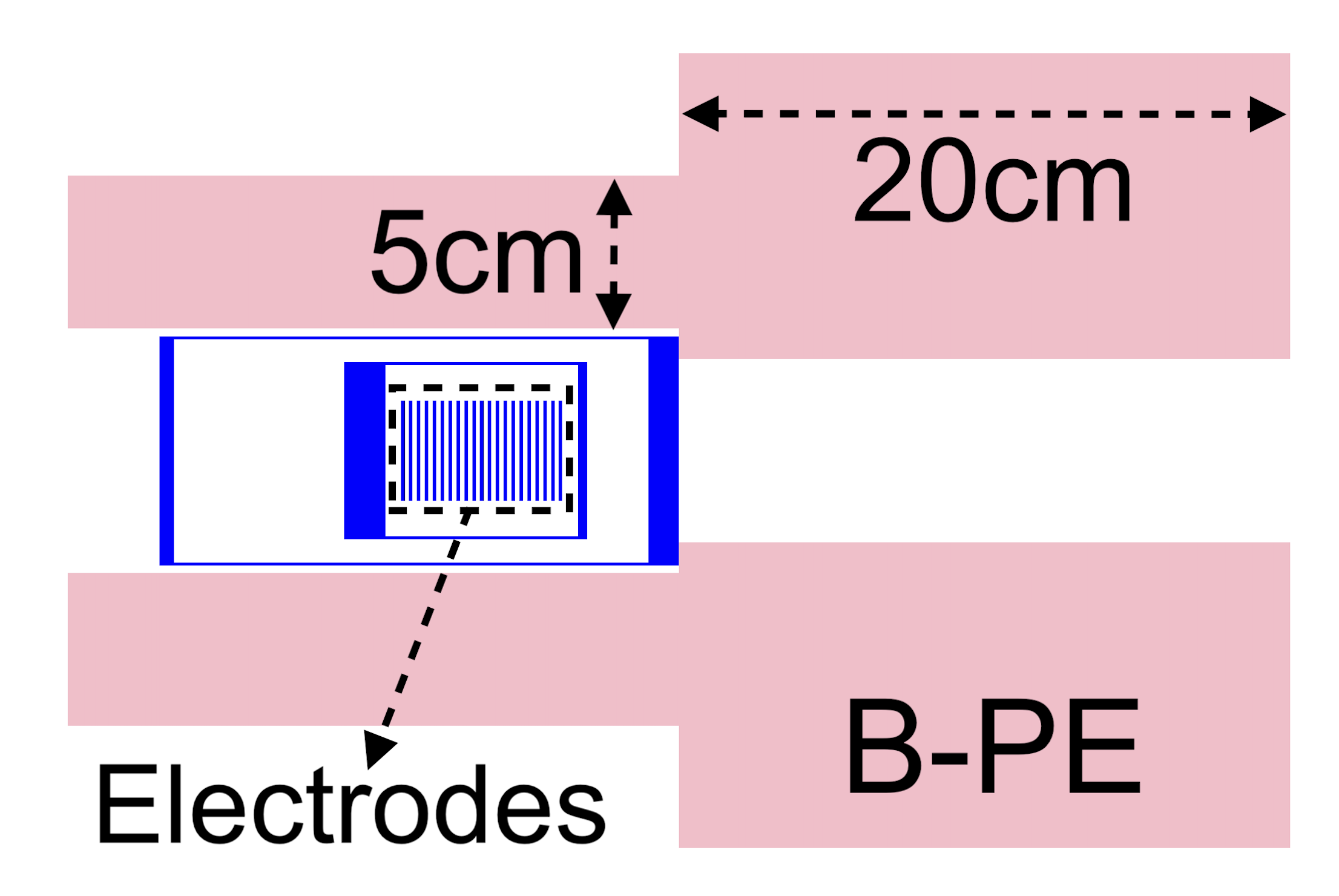}
                \caption{}
                \label{fig:coll_diagram}
            \end{subfigure}
            \caption{\add{(a)~Experimental setup and (b)~horizontal cross-section of the OpenMC model for the collimator tests with a DTn generator. The collimation is built with 5\% borated polyethylene (B-PE) blocks. In (b), $\mathrm{U_3 O_8}$ coatings are on both sides of each  electrodes, but too thin to be visible.}}
            \label{fig:U238_coll_diagram}
        \end{figure}

        To inform the SPARC design, primarily via validation of neutronics models, a bunker and collimation is built from 5\%-borated polyethylene (B-PE) bricks, as seen in \cref{fig:U238_coll_diagram}. Each brick has approximate dimensions $\SI{5}{cm} \times \SI{10}{cm} \times \SI{20}{cm}$, resulting in a square collimator of side length ${\sim}\SI{5}{cm}$ and depth ${\sim}\SI{20}{cm}$. The distance from face of the \ufc to the \dtng's axis of symmetry is fixed at ${\sim}\SI{45}{cm}$, while the translational distance, parallel to the axis, is scanned to bring the \dtng target plane in and out of the \ufc's field of view (FOV). 
        
        The resulting PHS and total count rates are shown in \cref{fig:coll_phs_openmc} for three translational offsets: $d = 0, 5$, and $\SI{10}{cm}$. Note that for these offset values, the total distance between the target plane and detector face only changes by ${\sim}\SI{1}{cm}$, or ${\sim}2\%$. However, the uncertainty in these offset values is (unfortunately) relatively large, $\Delta d \sim \SI{2}{cm}$, even though the uncertainty from counting statistics is too small to see on the plot. \add{In future work, both a finer scan in offset position and a mirrored translation (i.e. a negative offset) would be valuable for comparison with simulation, perhaps sacrificing uncertainty in counting statistics (e.g. shorter irradiation durations) for additional data points.}


        \renewcommand{\tempwidth}{0.49\linewidth}
        \begin{figure}[h!]
            \centering
            \begin{subfigure}{\tempwidth}
                \centering
                \includegraphics[width=\linewidth]{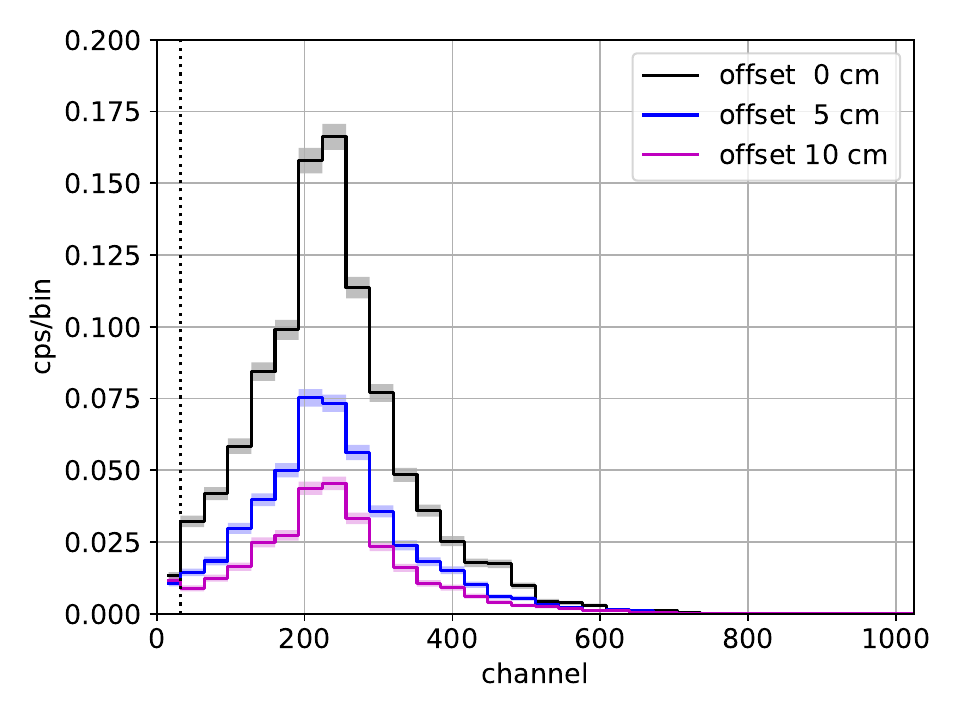}
                \caption{}
                \label{fig:coll_phs}
            \end{subfigure}
            \begin{subfigure}{\tempwidth}
                \centering
                \includegraphics[width=\linewidth]{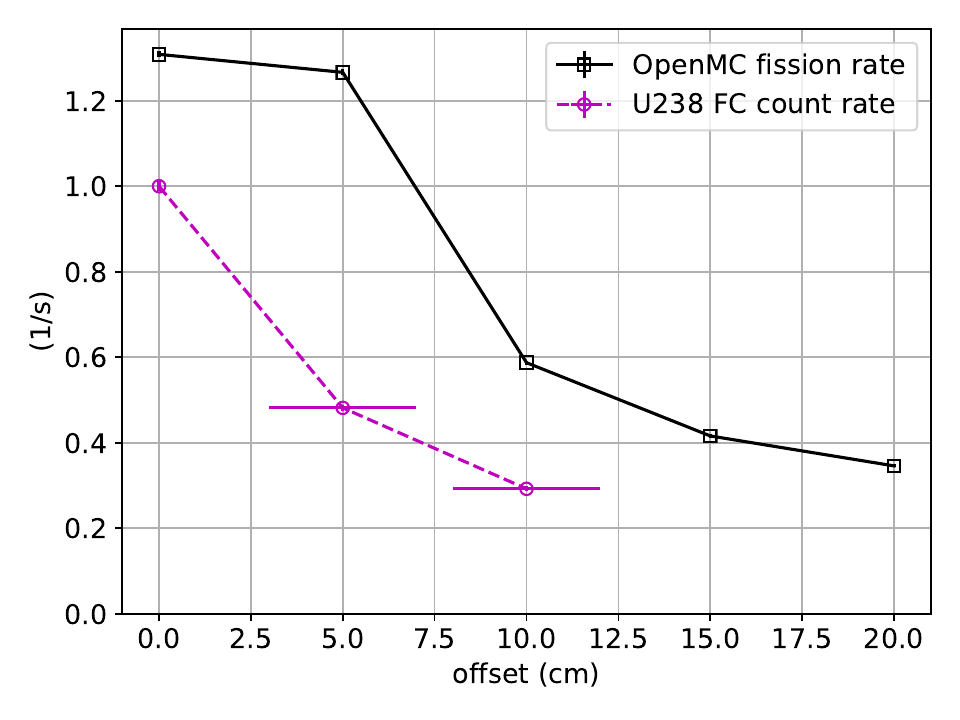}
                \caption{}
                \label{fig:coll_openmc}
            \end{subfigure}
            \caption{(a)~Pulse height spectra and (b)~total count rates - integrated above the dotted threshold in (a) - for the \ufc collimation study. Horizontal offsets from the \dtng target plane are scanned from ${\sim}\SI{0-10}{cm}$ in experiment and $\SI{0-20}{cm}$ in OpenMC neutronics simulations, the latter assuming a nominal \dtng source rate of $\SI{10^{8}}{n/s}$. Error bars are plotted, although some are too small to see.}
            \label{fig:coll_phs_openmc}
        \end{figure}

        \add{
        The OpenMC neutronics simulation code \cite{ROMANO201590} is used to complement this experimental analysis. The \dtng is simulated as a point source with experimentally measured spatial and spectral emissivity profiles presented in \cite{TSakabe_2024}. Polyethylene bricks with 5\% boron content (B-PE) are modeled with vendor-specified compositions and measured dimensions and densities ($\SI{1.03\pm0.02}{g/cm^3}$). Their configuration is reproduced from \cref{fig:collimator}, including the $\SI{5}{cm}$ square, $\SI{20}{cm}$ long collimator. The concrete walls and floors of the lab are also included, and (apart from the detector) the setup is simplified to include no other objects within the lab. 
        }

        \add{
        The \ufc is modeled in OpenMC via constructive solid geometry, with a cross-section shown in \cref{fig:coll_diagram}. The dimensions and compositions of the main components are preserved. The $\mathrm{U_3 O_8}$ coatings are present on both sides of each electrode, but they are too thin (${\sim}\SI{0.5}{\mu m}$) to be visible in the figure. Neutron flux energy spectra and fission rates in the coatings are tallied, the latter most directly related to count rates. Here, the total fission rate is scaled by a nominal \dtng source rate of ${\sim}\SI{10^8}{n/s}$.
        }

        \add{ We find relatively good agreement between simulation and experiment in \cref{fig:coll_phs_openmc}. As expected, the computed fission rates are slightly higher than the measured count rates because not every fission reaction results in a count.}
        Both the fission and count rates decrease with offset distance, yet there is some discrepancy in the ``decay length''. In OpenMC, the fission rate changes only minimally with a $d = \SI{5}{cm}$ offset because the DT point source is still within the \ufc's FOV. Then, the fission rate approximately halves when moving from 5 to $\SI{10}{cm}$. In experiment, a similar decrease in count rate is observed between 0 and $\SI{5}{cm}$; this could be due to the uncertainty in the \ufc placement \add{with respect to \dtng target plane}, meaning the ``true'' offset is ${>}\SI{5}{cm}$. In fact, the profiles in \cref{fig:coll_openmc} would almost overlap if the empirical data were shifted by $\Delta d = {+}\SI{5}{cm}$ and the simulation data were scaled by a slightly lower source rate. For SPARC, these results imply that a longer collimator would help to narrow the \ufc's FOV, and thicker side shielding could help to reduce indirect neutron measurements. \add{In addition, an \emph{in situ} calibration, with improved \dtng positioning and finer spatial steps, could help to better define the FOV.}        

    \subsection{Magnetic field sensitivity}\label{sec:expt-bfield}


        For the SPARC tokamak, the poloidal magnetic field coils will generate a primarily vertical ``stray'' field at the location of the \ufcs, with maximum magnitude ${\sim}\SI{250}{G}$ or $\SI{25}{mT}$. Thus, it is important to verify the robustness of the \ufc~- and other neutron flux monitors \cite{Wang2026_BF3} - to an externally applied \bfield. The parallel plate structure of the \ufc and the close spacing between plates $O(\SI{1}{mm})$ is expected to mitigate this effect. A simple analysis of electron trajectories, in a constant electric field (between plates) and a transverse \bfield, finds drift orbits with radii ${>}\SI{1}{cm}$, i.e. much larger than the cathode to anode separation. \add{This suggests that the \ufc design should be robust to a strong \bfield with little-to-no signal degradation.}

\begin{figure}[h!]
            \centering
            \includegraphics[width=0.5\linewidth]{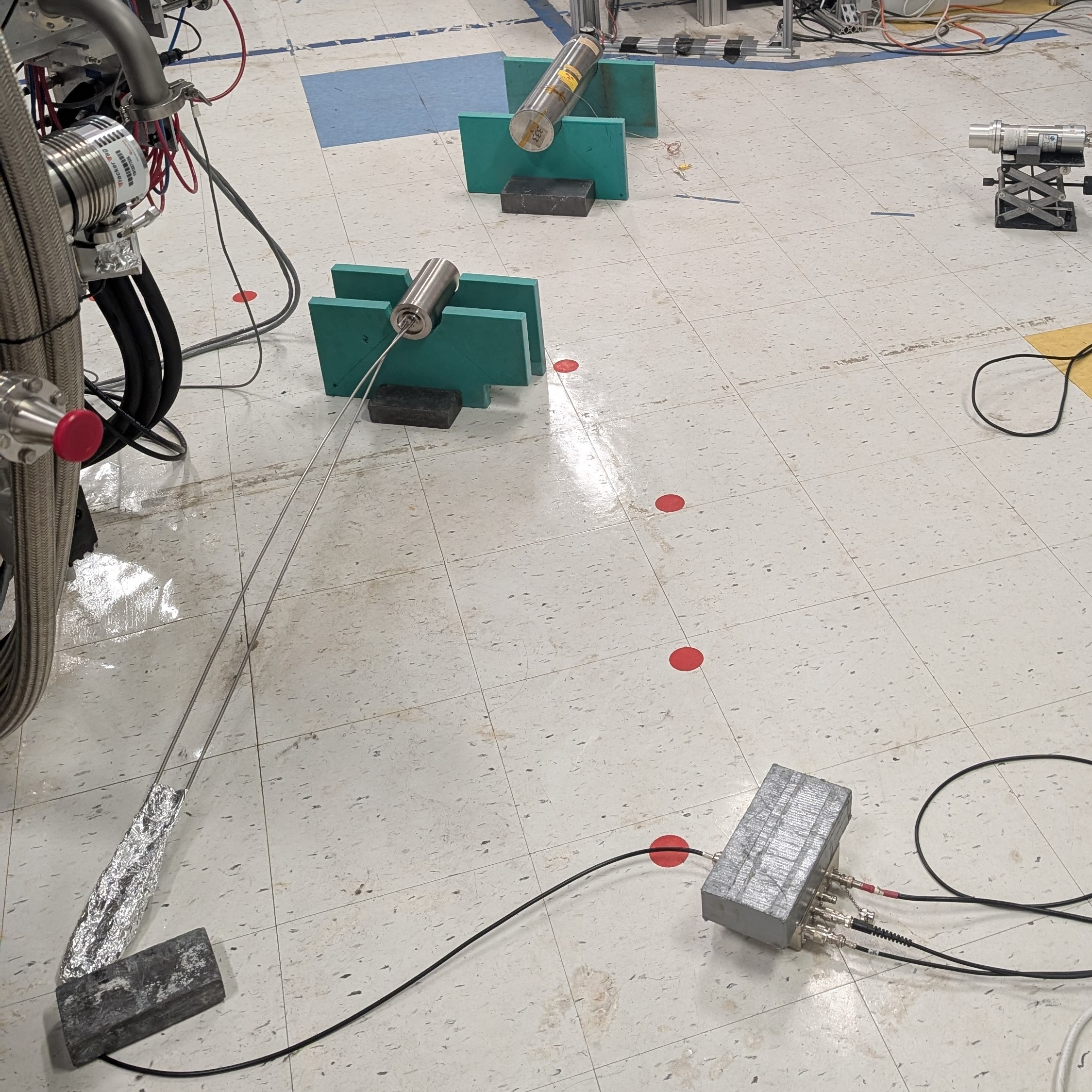}
            \caption{Experimental setup of the \ufc within the radiation field of the \dtng (top center) and the nominal 100~G line (red dots) of a cyclotron (at left, out of frame). Note that the \ufc leads and preamplifier are secured / weighed down by lead bricks (bottom). The DLOS reference detector is seen at top right.}
            \label{fig:cyclotron}
        \end{figure}

        This experiment is relatively challenging to recreate outside of a true tokamak environment, i.e. to place the \ufc within sufficiently strong radiation and magnetic fields simultaneously. Yet this could be achieved using the stray dipole \bfield of a proton cyclotron, which can be partially seen in the experimental setup of \cref{fig:cyclotron}. The cyclotron magnets are operated at half and full current, ${\sim}\SI{65}{A}$ and $\SI{130}{A}$, respectively, resulting in Hall sensor measurements of ${\sim}\SI{22}{mT}$ and $\SI{42}{mT}$ near the magnets. The nominal $\SI{10}{mT} = \SI{100}{G}$ boundary (for full current) is shown by the red dots in \cref{fig:cyclotron}. 

        The \dtng must be placed as close as possible to the detector (which sits within the $\SI{100}{G}$ line), but far enough away from the magnets so as to mitigate the impact of the stray \bfield on the accelerator-based source. \add{Unfortunately, in this experimental setup, the finite length of the \dtng power cable forces a tradeoff in source-detector distance and NG angular orientation; here, we prioritize minimizing distance and maximizing counts. Therefore,} as seen in \cref{fig:cyclotron}, the ``optimal'' configuration, given all constraints, happens to be the $\SI{0}{degree}$ orientation, such that the axis of the \dtng points radially toward the cyclotron. Thus, the DT neutron measurements are not made at $\SI{90}{deg}$, as in the previous sections' experiments. Nevertheless, here we are primarily interested in relative comparisons, which do not require a precise neutron energy.

        The \ufc is placed at two distances from the \dtng, and three measurements are made. First, the separation between the \dtng target plane and front face of the \ufc is set at ${\sim}\SI{21}{cm}$, with the detector entirely outside of the $\SI{100}{G}$ line. A hand-held \bfield meter registers strengths spanning $B \approx \SI{1-2}{mT}$ at the front and back of the detector, respectively, i.e. farther from and closer to the cyclotron. Note that the \bfield map outside of the cyclotron is not well characterized, so the true field is a mix of components transverse and parallel \add{to} the axis of \ufc. The second separation is ${\sim}\SI{83}{cm}$, with \bfield measurements $B \approx \SI{4-7}{mT}$ at half current. The third and final measurement is at the same separation distance but at full cyclotron current, with $B \approx \SI{8-14}{mT}$. 

\begin{figure}[h!]
            \centering
            \includegraphics[width=0.5\linewidth]{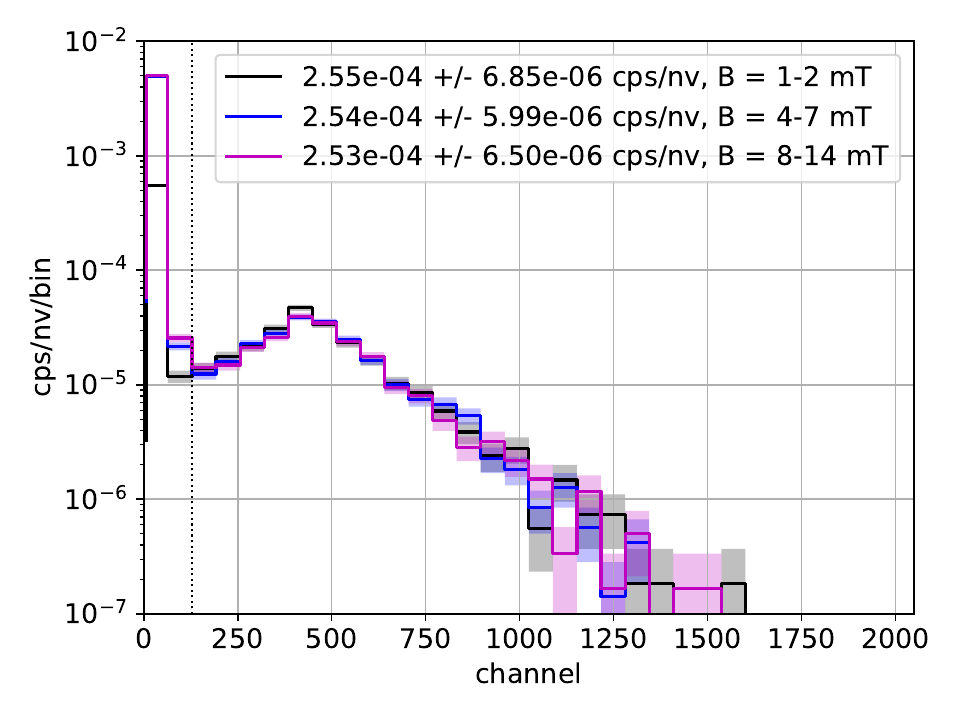}
            \caption{Pulse height spectrum (PHS) of the \ufc with shaded uncertainties for three magnetic field strengths $B = \SI{1-2}{mT}$ (black), $\SI{4-7}{mT}$ (blue), and $\SI{8-14}{mT}$ (magenta). The count rate (cps) is normalized by the nominal neutron flux (nv = n/cm$^2$/s), assuming a total \dtng rate of $\SI{10^8}{n/s}$ and isotropic emission over $\SI{4\pi}{sr}$. The total normalized rate with uncertainties, summed above a threshold (channel 128, vertical dotted line), is provided. Note the logarithmic vertical scale.}
            \label{fig:bfield}
        \end{figure}

        PHS are shown for the three measurements in \cref{fig:bfield}. Note that because the distance between the \ufc and \dtng is varied, each measurement is normalized to the effective neutron flux - as discussed in \cref{sec:expt-lin} - assuming a nominal DT neutron source rate of $\SI{10^8}{n/s}$. We see that the PHS overlap within error bars, at least for channels above 128. Integrating above this threshold gives the total count rate, excluding background and noise, which is provided in the legend of \cref{fig:bfield}. These values also agree well within uncertainties and match the vendor's specification.
        Even though SPARC's true \bfield strength of ${\sim}\SI{25}{mT}$ could not be achieved in these experiments, these results suggest that the increase from ${\sim}\SI{14}{mT}$ will have minimal impact on \ufc performance.

    \section{Summary}\label{sec:summary}

    The SPARC tokamak will deploy a variety of neutron flux monitors covering the entire dynamic range of an in-situ calibration source's rate ${\sim}\SI{10^8}{n/s}$ to the highest performing ``primary reference discharge'' plasma's ${>}\SI{10^{19}}{n/s}$. Along with boron-based gamma-compensated ionization chambers \cite{Raj2026}, a set of \U{238}-based fission chambers (FCs) are planned to cover the upper end of this range ${>}\SI{10^{15}}{n/s}$. The fission cross-section of \U{238} (see \cref{fig:cx_zoom}) is well suited to measure ``direct'' fast fusion neutrons from both DD and DT reactions. In addition, the unique parallel plate structure of this \ufc model (see \cref{fig:U238_diagram_openmc}) is meant to efficiently capture neutrons impinging normal to the plates. Effective collimation (see \cref{fig:U238_coll_diagram}) is key to prioritizing these unscattered neutrons. Experiments with 5\% borated polyethylene collimation and shielding showed a ${>}50\%$ reduction in count rate as the DT neutron source moved out of the \ufc's field of view (see \cref{fig:coll_phs_openmc}). Perhaps more importantly, good agreement was found between measured count rates and computed fission rates from OpenMC, thereby providing confidence in design decisions made with neutronics tools for the final SPARC design.

    A vendor-provided efficiency of $\SI{2.264\times 10^{-4}}{cps/nv}$ was corroborated by tests with both DD and DT neutron generators (see \cref{fig:U238_DD_DT}), at least within a factor of 2. This is reasonable given the uncertainties in the total DD and \dtng source rates, which were also found to be within a factor of 2 of the expected nominal rates (see \cref{fig:dd_dt_lin_phs}). Evidence suggests an increased sensitivity to DT compared to DD neutrons, as expected from the fission cross-section; however, this is difficult to confirm outside of experimental uncertainties. Good linearity with neutron rate was also confirmed and found to be relatively insensitive to the choice of threshold to exclude background signal and electronic noise. Additional discrimination of signals, e.g. between alpha decay and the fission fragments, is achievable from the pulse height spectrum (see \cref{fig:dt_lin_phs}); this should be doable in SPARC, although not strictly necessary for operations.

    A novel and challenging test of this \ufc involved characterizing the detector's response within a strong background magnetic field. In the SPARC setting, a stray \bfield up to ${\sim}\SI{25}{mT}$ is expected at the location of the \ufc and other neutron flux monitors. In this work, we evaluated the detector's performance up to $\SI{14}{mT}$ within the dipole field of a cyclotron (see \cref{fig:cyclotron}). \add{No significant change in count rate was identified when scaling from $\SI{1-10}{mT}$ (see \cref{fig:bfield}), pointing to an expected robustness for \ufc operation in the SPARC environment.}
    

    In a similar way, little change was observed when swapping short $\SI{1}{m}$ cables for long $\SI{30}{m}$ cables (see \cref{fig:wf}) nor when comparing fast $O(\SI{100}{ns})$ vs slow $O(\SI{10}{\mu s})$ preamplifiers. However, aluminum foil was used to wrap the connection between the \ufc leads and preamplifier cable (see \cref{fig:preamp_PS}) in order to shield out an unknown source of electronic noise. In the SPARC setting, the ${\sim}\SI{30}{m}$ cables will be required to take the signal from the tokamak to diagnostic hall; additionally, various sources of electromagnetic interference are anticipated. Here, we have identified multiple workable solutions for SPARC's \ufcs.

    \section*{Acknowledgments}

    The authors gratefully acknowledge funding from the MIT FUSAR program and Commonwealth Fusion Systems. Thanks to A.~Chouinard for assistance with the cyclotron magnetic field testing.
    \section*{References}
        \bibliographystyle{unsrt}
        \bibliography{bib}

@article{Ball2026,
doi = {10.1088/1741-4326/ae0faa},
url = {https://doi.org/10.1088/1741-4326/ae0faa},
year = {2025},
month = {nov},
publisher = {IOP Publishing},
volume = {66},
number = {1},
pages = {016010},
author = {Ball, J.L. and Mackie, S. and van de Lindt, J.G. and Morrissey, W. and Perevalov, A. and Short, Z.D. and Schwartz, N.R. and Koeth, T. and Beaudoin, B.L. and Romero-Talamás, C.A. and Rice, J.E. and Tinguely, R.A.},
title = {{Measurements of fusion yield on the Centrifugal Mirror Fusion Experiment}},
journal = {Nuclear Fusion}
}

@article{Batistoni2018,
doi = {10.1088/1741-4326/aad4c1},
url = {https://doi.org/10.1088/1741-4326/aad4c1},
year = {2018},
month = {aug},
publisher = {IOP Publishing},
volume = {58},
number = {10},
pages = {106016},
author = {Batistoni, P. and Popovichev, S. and Ghani, Z. and Cufar, A. and Giacomelli, L. and Hawkins, P. and Keogh, K. and Jednorog, S. and Laszynska, E. and Loreti, S. and Peacock, A. and Pillon, M. and Price, R. and Reed, A. and Rigamonti, D. and Stephens, J. and Bielecki, J. and Conroy, S. and Dankowski, J. and Krasilnikov, V. and JET contributors},
title = {{14 MeV calibration of JET neutron detectors—phase 2: in-vessel calibration}},
journal = {Nuclear Fusion}
}

@article{Panontin2026,
    author = {E. Panontin and others},
    title = {{Synthetic model of gamma-ray emission during DT experiments on the SPARC tokamak}},
    journal = {Nuclear Fusion},
    year = {2026},
    note = {In progress.}
}

@article{Taieb2016,
title = {A new fission chamber dedicated to Prompt Fission Neutron Spectra measurements},
journal = {Nuclear Instruments and Methods in Physics Research Section A: Accelerators, Spectrometers, Detectors and Associated Equipment},
volume = {833},
pages = {1-7},
year = {2016},
issn = {0168-9002},
doi = {https://doi.org/10.1016/j.nima.2016.06.137},
url = {https://www.sciencedirect.com/science/article/pii/S0168900216307070},
author = {J. Taieb and B. Laurent and G. Bélier and A. Sardet and C. Varignon}
}

@article{Wang2026_BF3,
    author = {X. Wang and others},
    title = {{Characterization of a BF$_3$ Proportional Counter for Neutron Measurements in the SPARC Tokamak}},
    journal = {Fusion Engineering and Design},
    year = {2026},
    note = {In progress.},
}

@article{Raj2026,
    author = {P. Raj and others},
    title = {{Final design of the SPARC neutron flux monitors for fusion power diagnostics}},
    journal = {Fusion Engineering and Design},
    year = {2026},
    note = {In progress.}
}

@article{Jarvis1990,
    author = {Jarvis, O. N. and Sadler, G. and van Belle, P. and Elevant, T.},
    title = {In‐vessel calibration of the JET neutron monitors using a 252Cf neutron source: Difficulties experienced},
    journal = {Review of Scientific Instruments},
    volume = {61},
    number = {10},
    pages = {3172-3174},
    year = {1990},
    month = {10},
    issn = {0034-6748},
    doi = {10.1063/1.1141677},
    url = {https://doi.org/10.1063/1.1141677},
    eprint = {https://pubs.aip.org/aip/rsi/article-pdf/61/10/3172/19036136/3172_1_online.pdf},
}

@article{Nieschmidt1985,
    author = {Nieschmidt, E. B. and England, A. C. and Hendel, H. W. and Hillis, D. L. and Isaacson, J. A. and Ku, L. P. and Tsang, F. Y.},
    title = {Effects of neutron energy spectrum on the efficiency calibration of epithermal neutron detectors},
    journal = {Review of Scientific Instruments},
    volume = {56},
    number = {5},
    pages = {1084-1086},
    year = {1985},
    month = {05},
    issn = {0034-6748},
    doi = {10.1063/1.1138230},
    url = {https://doi.org/10.1063/1.1138230},
    eprint = {https://pubs.aip.org/aip/rsi/article-pdf/56/5/1084/19116559/1084_1_online.pdf},
}

@article{Fiore1995,
    author = {Fiore, C. L. and Boivin, R. L.},
    title = {{Performance of the neutron diagnostic system for Alcator C‐Mod}},
    journal = {Review of Scientific Instruments},
    volume = {66},
    number = {1},
    pages = {945-947},
    year = {1995},
    month = {01},
    issn = {0034-6748},
    doi = {10.1063/1.1146215},
    url = {https://doi.org/10.1063/1.1146215},
    eprint = {https://pubs.aip.org/aip/rsi/article-pdf/66/1/945/19296692/945_1_online.pdf},
}

@article{Jarvis1994,
doi = {10.1088/0741-3335/36/2/002},
url = {https://doi.org/10.1088/0741-3335/36/2/002},
year = {1994},
month = {feb},
publisher = {},
volume = {36},
number = {2},
pages = {209},
author = {O N Jarvis},
title = {Neutron measurement techniques for tokamak plasmas},
journal = {Plasma Physics and Controlled Fusion}
}

@article{JENDL32,
author = {Tsuneo NAKAGAWA and Keiichi SHIBATA and Satoshi CHIBA and Tokio FUKAHORI and Yutaka NAKAJIMA and Yasuyuki KIKUCHI and Toshihiko KAWANO and Yukinori KANDA and Takaaki OHSAWA and Hiroyuki MATSUNOBU and Masayoshi KAWAI and Atsushi ZUKERAN and Takashi WATANABE and Sin-iti IGARASI and Kazuaki KOSAKO and Tetsuo ASAMI},
title = {Japanese Evaluated Nuclear Data Library Version 3 Revision-2: JENDL-3.2},
journal = {Journal of Nuclear Science and Technology},
volume = {32},
number = {12},
pages = {1259--1271},
year = {1995},
publisher = {Taylor \& Francis},
doi = {10.1080/18811248.1995.9731849},
URL = { 
        https://doi.org/10.1080/18811248.1995.9731849
},
eprint = { 
        https://doi.org/10.1080/18811248.1995.9731849
}
}

@article{JENDL33,
author = {Keiichi SHIBATA and Toshihiko KAWANO and Tsuneo NAKAGAWA and Osamu IWAMOTO and Jun-ichi KATAKURA and Tokio FUKAHORI and Satoshi CHIBA and Akira HASEGAWA and Toru MURATA and Hiroyuki MATSUNOBU and Takaaki OHSAWA and Yutaka NAKAJIMA and Tadashi YOSHIDA and Atsushi ZUKERAN and Masayoshi KAWAI and Mamoru BABA and Makoto ISHIKAWA and Tetsuo ASAMI and Takashi WATANABE and Yukinobu WATANABE and Masayuki IGASHIRA and Nobuhiro YAMAMURO and Hideo KITAZAWA and Naoki YAMANO and Hideki TAKANO},
title = {Japanese Evaluated Nuclear Data Library Version 3 Revision-3: JENDL-3.3},
journal = {Journal of Nuclear Science and Technology},
volume = {39},
number = {11},
pages = {1125--1136},
year = {2002},
publisher = {Taylor \& Francis},
doi = {10.1080/18811248.2002.9715303},
URL = { 
        https://doi.org/10.1080/18811248.2002.9715303
},
eprint = { 
        https://doi.org/10.1080/18811248.2002.9715303
}
}

@article{JENDL4,
author = {Keiichi SHIBATA and Osamu IWAMOTO and Tsuneo NAKAGAWA and Nobuyuki IWAMOTO and Akira ICHIHARA and Satoshi KUNIEDA and Satoshi CHIBA and Kazuyoshi FURUTAKA and Naohiko OTUKA and Takaaki OHSAWA and Toru MURATA and Hiroyuki MATSUNOBU and Atsushi ZUKERAN and So KAMADA and Jun-ichi KATAKURA},
title = {JENDL-4.0: A New Library for Nuclear Science and Engineering},
journal = {Journal of Nuclear Science and Technology},
volume = {48},
number = {1},
pages = {1--30},
year = {2011},
publisher = {Taylor \& Francis},
doi = {10.1080/18811248.2011.9711675},
URL = { 
        https://doi.org/10.1080/18811248.2011.9711675
},
eprint = { 
        https://doi.org/10.1080/18811248.2011.9711675
}
}

@article{JENDL5,
author = {Osamu Iwamoto and Nobuyuki Iwamoto and Satoshi Kunieda and Futoshi Minato and Shinsuke Nakayama and Yutaka Abe and Kohsuke Tsubakihara and Shin Okumura and Chikako Ishizuka and Tadashi Yoshida and Satoshi Chiba and Naohiko Otuka and Jean-Christophe Sublet and Hiroki Iwamoto and Kazuyoshi Yamamoto and Yasunobu Nagaya and Kenichi Tada and Chikara Konno and Norihiro Matsuda and Kenji Yokoyama and Hiroshi Taninaka and Akito Oizumi and Masahiro Fukushima and Shoichiro Okita and Go Chiba and Satoshi Sato and Masayuki Ohta and Saerom Kwon},
title = {Japanese evaluated nuclear data library version 5: JENDL-5},
journal = {Journal of Nuclear Science and Technology},
volume = {60},
number = {1},
pages = {1--60},
year = {2023},
publisher = {Taylor \& Francis},
doi = {10.1080/00223131.2022.2141903},
URL = { 
        https://doi.org/10.1080/00223131.2022.2141903
},
eprint = { 
        https://doi.org/10.1080/00223131.2022.2141903
}
}

@article{Ball2024,
    author = {Ball, J. L. and Panontin, E. and Mackie, S. and Tinguely, R. A. and Raj, P.},
    title = {Evaluating deuterated-xylene for use as a fusion neutron spectrometer},
    journal = {Review of Scientific Instruments},
    volume = {95},
    number = {12},
    pages = {123514},
    year = {2024},
    month = {12},
    issn = {0034-6748},
    doi = {10.1063/5.0219490},
    url = {https://doi.org/10.1063/5.0219490},
    eprint = {https://pubs.aip.org/aip/rsi/article-pdf/doi/10.1063/5.0219490/20304345/123514_1_5.0219490.pdf},
}

@article{Becchetti2016,
title = {Deuterated-xylene (xylene-d10; EJ301D): A new, improved deuterated liquid scintillator for neutron energy measurements without time-of-flight},
journal = {Nuclear Instruments and Methods in Physics Research Section A: Accelerators, Spectrometers, Detectors and Associated Equipment},
volume = {820},
pages = {112-120},
year = {2016},
issn = {0168-9002},
doi = {https://doi.org/10.1016/j.nima.2016.02.058},
url = {https://www.sciencedirect.com/science/article/pii/S0168900216002345},
author = {F.D. Becchetti and R.S. Raymond and R.O. Torres-Isea and A. {Di Fulvio} and S.D. Clarke and S.A. Pozzi and M. Febbraro}
}

@article{Raj2024,
    author = {Raj, P. and Ball, J. L. and Carmichael, J. and Frenje, J. A. and Gocht, R. and Gorini, G. and Holmes, I. and Johnson, M. Gatu and Kennedy, R. and Mackie, S. and Nocente, M. and Panontin, E. and Petruzzo, M. and Rebai, M. and Reinke, M. and Rice, J. and Rigamonti, D. and Rosa, M. Dalla and Saltos, A. A. and Tardocchi, M. and Tinguely, R. A. and Wang, X.},
    title = {{Overview of the neutron diagnostic systems for the SPARC tokamak}},
    journal = {Review of Scientific Instruments},
    volume = {95},
    number = {10},
    pages = {103507},
    year = {2024},
    month = {10},
    issn = {0034-6748},
    doi = {10.1063/5.0219538},
    url = {https://doi.org/10.1063/5.0219538},
    eprint = {https://pubs.aip.org/aip/rsi/article-pdf/doi/10.1063/5.0219538/20213486/103507_1_5.0219538.pdf},
}

@article{Creely2020,
author = {Creely, A. J. and Greenwald, M. J. and Ballinger, S. B. and Brunner, D. and Canik, J. and Doody, J. and F{\"{u}}l{\"{o}}p, T. and Garnier, D. T. and Granetz, R. and Gray, T. K. and Holland, C. and Howard, N. T. and Hughes, J. W. and Irby, J. H. and Izzo, V. A. and Kramer, G. J. and Kuang, A. Q. and LaBombard, B. and Lin, Y. and Lipschultz, B. and Logan, N. C. and Lore, J. D. and Marmar, E. S. and Montes, K. and Mumgaard, R. T. and Paz-Soldan, C. and Rea, C. and Reinke, M. L. and Rodriguez-Fernandez, P. and S{\"{a}}rkim{\"{a}}ki, K. and Sciortino, F. and Scott, S. D. and Snicker, A. and Snyder, P. B. and Sorbom, B. N. and Sweeney, R. and Tinguely, R. A. and Tolman, E. A. and Umansky, M. and Vallhagen, O. and Varje, J. and Whyte, D. G. and Wright, J. C. and Wukitch, S. J. and Zhu, J.},
doi = {10.1017/S0022377820001257},
issn = {0022-3778},
journal = {Journal of Plasma Physics},
month = {oct},
number = {5},
pages = {865860502},
publisher = {Cambridge University Press},
title = {{Overview of the SPARC tokamak}},
url = {https://www.cambridge.org/core/product/identifier/S0022377820001257/type/journal_article},
volume = {86},
year = {2020}
}

@article{ROMANO201590,
title = {OpenMC: A state-of-the-art Monte Carlo code for research and development},
journal = {Annals of Nuclear Energy},
volume = {82},
pages = {90-97},
year = {2015},
issn = {0306-4549},
doi = {https://doi.org/10.1016/j.anucene.2014.07.048},
author = {Paul K. Romano and others}
}

@article{TSakabe_2024,
  author={Sakabe, T. and Edwards, E. and Lanzrath, A. and Goles, N. and Ball, J. and Mackie, S. and Segantin, S. and Mukai, K. and Yagi, J. and Tinguely, R. A. and Woller, K. B.},
  journal={IEEE Transactions on Plasma Science}, 
  title={{Neutron Energy Distribution and Energy-Corrected Spatial Distribution Around a Sealed-Tube DT Fusion Neutron Generator}}, 
  year={2024},
  volume={52},
  number={9},
  pages={3871-3877},
  doi={10.1109/TPS.2024.3384234}}

\end{document}